\documentclass[prl,aps,twocolumn,groupedaddress,floats,showpacs,final,superscriptaddress]{revtex4-1}
\usepackage{leading}
\usepackage[latin3]{inputenc}
\usepackage[makeroom]{cancel}
\usepackage{graphicx}
\usepackage{amsmath}
\usepackage{physics}
\usepackage{amsfonts}
\usepackage{babel}
\usepackage{amssymb}

\usepackage{color}
\usepackage[left=2cm,right=2cm,top=2cm,bottom=2cm]{geometry}
\usepackage{graphicx} 
\usepackage{dcolumn}
\usepackage{bm}
\usepackage{simplewick}
\usepackage{array}
\usepackage{appendix}
\usepackage{nicefrac}
\usepackage{chngcntr}
\usepackage[left=2cm,right=2cm,top=2cm,bottom=2cm]{geometry}
\usepackage[normalem]{ulem}

\RequirePackage[
   hyperindex,colorlinks,bookmarksnumbered,
   plainpages=true,pdfstartview=FitH]{hyperref}
\hypersetup{linkcolor=blue,urlcolor=blue,citecolor=blue}
\definecolor{purple}{rgb}{0.5,0,0.6}

\renewcommand{\emph}[1]{\textit{#1}}
\definecolor{darkblue}{rgb}{0,0,0.5}
\definecolor{darkgreen}{rgb}{0,0.5,0}
\definecolor{darkred}{rgb}{.7,0,0}
\definecolor{purple}{rgb}{0.5,0,0.6}
\definecolor{orange}{rgb}{1,0.5,0}
\definecolor{grey}{rgb}{.6,.6,.6}
\definecolor{lightpink}{rgb}{1,0.7,0.75}
\definecolor{pink}{rgb}{1,0.4,0.58}
\definecolor{deeppink}{rgb}{1,0.08,0.58}

\newcommand{\beq}{\begin{equation}}
\newcommand{\eeq}{\end{equation}}
\newcommand{\bea}{\begin{eqnarray}}
\newcommand{\eea}{\end{eqnarray}}
\newcommand{\nn}{\nonumber}

\begin{document}
\title{
Quantum thermal transport in the charged Sachdev-Ye-Kitaev model:\\ Thermoelectric Coulomb blockade
 }
 
\author{Andrei I. Pavlov}
\author{Mikhail N. Kiselev}
\affiliation{The Abdus Salam International Centre for Theoretical Physics (ICTP), Strada Costiera 11, I-34151 Trieste, Italy}
\begin{abstract}
We present a microscopic theory for quantum thermoelectric and heat transport in the 
Schwarzian regime of the Sachdev-Ye-Kitaev (SYK) model. As a charged fermion realization of the SYK model in nanostructures we assume a setup based on a quantum dot connected to the charge reservoirs through weak tunnel barriers. We analyze particle-hole symmetry breaking effects crucial for both Seebeck and Peltier coefficients. We show that the quantum charge and heat transport at low temperatures are defined by the interplay between elastic and inelastic processes such that the inelastic processes provide a leading contribution to the transport coefficients at the temperatures that are smaller compared to the charging energy. We demonstrate that both electric and thermal conductance obey a power law in temperature behavior, while thermoelectric, Seebeck, and Peltier coefficients are exponentially suppressed. This represents selective suppression of only nondiagonal transport coefficients. We discuss the validity of the Kelvin formula in the presence of a strong Coulomb blockade.

\end{abstract}

\maketitle

\paragraph*{Introduction. }
In recent years, the Sachdev-Ye-Kitaev (SYK) model \cite{Sachdev1993, Kitaev2015} has become popular in numerous studies due to its unique features. This model, which can be formulated in terms of Majorana fermions or conventional complex fermions (cSYK) in $(0+1)$ dimensions, has solvable nontrivial limits with absent quasiparticles, saturates the bound for quantum chaos \cite{Maldacena2016jhep, Jensen2016}, and is holographically connected to black holes with $AdS_2$ [$(1+1)$ dimensional anti-de-Sitter] horizons \cite{Engelsy2016}. Extensions of the model to the coupled cSYK clusters reproduce the benchmark properties of strange metals \cite{Patel2018, Patel2019}, such as linear in temperate resistivity \cite{Song2017} and thermal diffusivity \cite{Davison2017}, observed in cuprates \cite{Legros2018}, pnictides \cite{Nakajima2019}, and twisted bilayer graphene \cite{Cao2020}. The cSYK model allows a nontrivial analytic saddle-point solution if the dynamics of the model can be neglected. This solution possesses both conformal $SL(2,\mathbb{R})$ and gauge $U(1)$ symmetries. The low-energy limit of the SYK model is governed by the symmetry breaking mechanism of Goldstone reparametrization modes \cite{Bagrets2016, Maldacena2016}, known as the Schwarzian regime in theories of gravity in a nearly $AdS_2$ space-time \cite{Jensen2016, Maldacena2016p, Engelsy2016}. The cSYK model with charge density $\mathcal{Q}$ is shown to possess the residual zero-temperature entropy $S$ per particle proportional to the parameter controlling particle-hole asymmetry $\mathcal{E}$ in the system, $\frac{\partial S}{\partial \mathcal{Q}}=2\pi \mathcal{E}$ \cite{Georges2001}, and identified as the Bekenstein-Hawking entropy of the charged black hole \cite{Sachdev2010prl, Sachdev2010, Sachdev2015}. The finite Bekenstein-Hawking entropy is present in both the conformal and Schwarzian regimes of the theory \cite{Gu2019}. Various experimental realizations of the SYK model are proposed in quantum gases \cite{Danshita2017}, Majorana wires \cite{Chew2017} and topological superconductor \cite{Pikulin2017} devices. Possible experimental realization of the cSYK model in irregularly shaped graphene flake quantum dots \cite{Chen2018, Can2019} opens possibilities for direct studies of thermoelectric transport properties of the model.

The thermoelectric transport through quantum dots is a subject of extensive theoretical \cite{Beenakker1992, Furusaki1995, Matveev2002, Turek2002} and experimental \cite{Staring1993, Dzurak1997, Scheibner2005, Potok2007} studies, as it opens broad possibilities for technological advancements in thermoelectric and microelectronic industries \cite{Dresselhaus1999, Zlatic2014, Benenti2017} and provides tools for better understanding of strongly correlated systems \cite{Zlatic1993, Kim2003}. Among all thermoelectric coefficients, the thermopower $\mathcal{S}$ is a subject of particular interest due to its high sensitivity to the particle-hole asymmetry of the system. The thermopower measurements allow probing of particle-hole asymmetry related effects and provide information about low-energy excitations in the system \cite{Matveev2002, Vavilov2005, Karki2019}. These properties of the thermopower make it a useful tool for capturing Fermi liquid--non-Fermi liquid transitions (FL-NFL) by accessing the NFL regime \cite{Nguyen2018, Nguyen2019}. 

Substantial progress was recently made in the entropy measurement of mesoscopic quantum systems \cite{Eriksson2011} of sizes up to a few particles \cite{Hartman2018, Sela2019}. One of the most interesting and promising developments in these fields is the study of the entropy via the thermoelectric properties of the system \cite{Yang2009, Chickering2013, Hou2012, Kleeorin2019}. This approach employs the fact that under certain conditions the thermopower can be regarded as the entropy per particle \cite{Shastry2008}. 
The thermoelectric transport coefficients relate the charge and heat current, $I_e$ and $I_h$, to applied voltage and temperature differences, $\Delta V$ and $\Delta T$, in the linear response regime as
\beq \label{currentI} \begin{matrix}
I_e=G \Delta V+G_T\Delta T, \\
I_h=G_T \Delta V+K\Delta T.
\end{matrix}\eeq
Here, $G$ is the electric conductance, $G_T$ is the thermoelectric coefficient, $\mathcal{S}=G_T/G$ is the thermopower, and $\kappa= K-G\mathcal{S}^2/T$ is the thermal conductance \cite{Onsager1931p1, Onsager1931p2, Kubala2008}.

The entropy and thermopower are connected by the Kelvin formula $\mathcal{S}$$=$$ \left(\frac{\partial S}{\partial N}\right)_{T, V} $, where $N$ is the number of particles \cite{Peterson2010} (we use units $e$$=$$\hbar$$=$$k_B$$=$$1$ for the electron charge, Planck's, and Boltzmann constants). This relation is an empirical approximation of the thermopower given in terms of the transport coefficients in Eq.(\ref{currentI}). It is considered to be applicable in the thermodynamic limit of the transport \cite{Mravlje2016}, but does not necessarily hold in a general case \cite{Deng2013, Karki2020}. While approximate in most cases, the Kelvin formula is shown to be exact in systems with the $SL(2,\mathbb{R})$ symmetry \cite{Davison2017}. One of the most well-studied examples of the systems realizing this symmetry is the SYK model in the conformal regime \cite{Sachdev2015}.

The transport properties of the cSYK dots are extensively studied in the conformal regime \cite{Sachdev2015, Gnezdilov2018, Can2019, Cha2020}. It is shown that the system exhibits FL-NFL transitions governed by the parameters of the dot. In a recent study \cite{Kruchkov2019}, the authors examined the 
Schwarzian regime of the SYK model realized in a quantum dot with the addition of the finite charging energy, $E_C$, and argued that the relation between the entropy and the thermopower holds there as well, which opens possibilities for direct measurements of the residual Bekenstein-Hawking entropy in the proposed experimental realization of the cSYK model \cite{Chen2018, Can2019}. However, these studies only account for direct tunneling (also known as elastic tunneling \cite{Pustilnik2004}) in quantum transport, while the inelastic co-tunneling transport can significantly affect the transport properties of quantum systems \cite{Furusaki1995, Matveev2002}, including the transport through the cSYK quantum dot \cite{Altland2019}.

In the present \textit{Letter}, we examine the low-temperature limit of quantum transport through the cSYK quantum dot and address the violation of the relation between thermopower and entropy due to the dominant role of inelastic processes in the presence of the Coulomb blockade.


\paragraph*{Model. }
We consider a cSYK quantum dot (QD) coupled to two identical metallic leads, in a setup similar to \cite{Gnezdilov2018, Kruchkov2019}. Fermions of the cSYK dot and the lead are coupled via the tunneling term. 

The Hamiltonian of the cSYK dot with $i$$=$$1...N$ electronic orbitals represented by $N$ complex spinless fermions \cite{comment_spin} is given by
\beq \label{Hsyk} H_{\rm SYK}=\frac{1}{(2N)^{3/2}}\sum_{ijkl=1}^NJ_{ij;kl}c^{\dagger}_ic^{\dagger}_jc_kc_l
-\mu\sum_{i=1}^Nc^{\dagger}_ic_i,
\eeq
where $J_{i,j;k,l}$ is the random Gaussian interaction constant with zero mean value $\langle J_{ij;kl}\rangle=0$ and nonzero variance $\langle|J_{ij;kl}|^2\rangle=J^2$, and $\mu$ is the chemical potential of fermions inside the dot. 
Since the leads are identical, we consider QD coupled to the symmetric superposition of the electronic states (one ``effective" contact). The full Hamiltonian, which takes the lead into account, reads
\beq \label{FullH} H=H_{\rm SYK}+E^{(0)}_C\hat{n}^2+\sum_{q}\varepsilon_qa^{\dagger}_qa_q+\sum_i\left(\lambda_ic^{\dagger}_ia_q+H.c.\right), \eeq
where $\lambda_i$ is the random tunneling constant, and we assume that it is Gaussian with zero mean $\langle\lambda_i\rangle=0$ and nonzero variance $\langle|\lambda_i|^2\rangle=\lambda^2$. Operators $a, \, a^{\dagger}$ represent a symmetric combination of fermions in the leads with dispersion relation $\varepsilon_q$. $E^{(0)}_C$ is the charging energy and $\hat{n}=\sum_{i}^Nc_i^{\dagger}c_i$ is the charge on the dot \cite{comment_Ec}.\\
The average tunneling term $\left\langle \lambda_i \right\rangle=0$, so the direct tunneling does not contribute to the system's currents. Since both the elastic and inelastic cotunnelings are present, this situation resembles tunneling through a barrier that is randomly fluctuating in time \cite{Kiselev2009}. For an arbitrary tunneling constant variance $\lambda^2$, the fermion Green's function (GF) of the dot is effectively renormalized in the presence of the lead. In the conformal limit of the cSYK model \cite{Kitaev2015, Sachdev2015}, this renormalization of the Green's function was discussed in \cite{Gnezdilov2018, Kruchkov2019}.  
The physics of an isolated SYK dot in the low temperature regime, $J e^{-N/2}\ll T\ll J/[N\ln N]$ (we denote, further, $T^*\equiv J/[N\ln N]$) is defined by the Schwarzian action (derived in \cite{Maldacena2016, Bagrets2016}), which appears due to breaking of the conformal symmetry $SL(2,\mathbb{R})$ group with an additional contribution to the the cSYK model from breaking $U(1)$ symmetry (discussed in \cite{Altland2019}). The inequality $J e^{-N/2}\ll T$ in the limit of large $N$ ensures that the effects of mean-level spacing can be neglected \cite{Bagrets2016}; 
all further results are discussed within this assumption. The appearing Goldstone modes renormalize the saddle point solution for the conformal GF \cite{comment_tunneling}. 
We assume that the dot-lead coupling $\lambda$ is the smallest energy scale of the system ($\lambda\ll \textit{min}\lbrace T,T^*)\rbrace$). It allows consideration of the system in the vicinity of the original saddle point of the isolated cSYK dot, similar to \cite{Altland2019, Can2019}. This assumption is not related to the effects of breaking the particle-hole symmetry in the conformal and Schwarzian limits of the model, of which we are interested in the scope of this \textit{Letter}.\\
At temperatures above the charging energy, $E^{(0)}_C\ll T\ll J$, the effects of the Coulomb blockade are reduced and the direct tunneling dominates the transport properties. This case was studied in detail in \cite{Kruchkov2019}. In the present \textit{Letter}, we focus on transport properties at low-energy scales, where inelastic co-tunneling processes give a crucial contribution to the currents. So, we are interested in two energy scales, $T^*\ll T\ll\textit{min} \lbrace E^{(0)}_C,J\rbrace$ and $T\ll T^*\ll\textit{min} \lbrace E_C,J\rbrace $. The former corresponds to the conformal regime, while the latter is described by the Schwarzian physics. The relevance of the latter case was additionally addressed in \cite{Altland2019}, as the charging energy is effectively renormalized by the Goldstone modes, $E_C=E^{(0)}_C+\mathcal{K}$, with the additional contribution to the charging energy, $\mathcal{K}\sim T^*$.  It ensures that the effective charging energy is always $E_C> T^*$. We suppose further that $E_C$ includes this renormalization of the charging energy. \\
The thermoelectric properties of the system are expressed via the electric $I_e$ and heat $I_h$ currents, given in the weak-tunneling limit by the Fermi golden rule \cite{Zlatic2014},

{\color{black}
\begin{eqnarray}I_e&=&-2\pi \int_{-\infty}^{\infty}d\varepsilon \rho_a(\epsilon)\rho_c(\epsilon)\Delta f(\epsilon, T), \nonumber\\
I_h&=&-2\pi \int_{-\infty}^{\infty}d\varepsilon \varepsilon\rho_a(\epsilon)\rho_c(\epsilon)\Delta f(\epsilon, T),
\label{FGL} 
\end{eqnarray}}
where $\rho_a$ and $\rho_c$ are density of states (DOS) in the lead and the dot correspondingly, $f(\epsilon, T)$ is the Fermi distribution function at temperature $T$, $\Delta V$ is the applied voltage and $\Delta f(\epsilon, T)= f(\epsilon +\Delta V, T+\Delta T)-f(\epsilon, T)$. The free fermion DOS in metal has weak energy dependence around the Fermi level, so it can be put to a constant $\rho_a=(2\pi v_F)^{-1}$ \cite{Gnezdilov2018, Can2019, Kruchkov2019}, where $v_F$ is the Fermi velocity. 

Equations (\ref{currentI}) and (\ref{FGL}) allow us to find the system's electric conductance $G$, thermoelectric coefficient $G_T$, and thermal conductance $\kappa$ (see Supplemental Material \cite{supplemental}). Following the approach of \cite{Furusaki1995, Matveev2002}, we express these entities through the $T$ matrix in the Matsubara representation $\mathcal{T}$ \cite{Kim2003} (the $T$ matrix is supposed to be momentum independent, which is the case for short-range interactions), 
{\color{black}
\begin{eqnarray} 
\label{Gformula} 
&&G=\frac{1}{2v_F}\int_{-\infty}^{\infty} dt
\frac{1}{\cosh\left(\pi T t\right) }\mathcal{T}\left(\frac{1}{2T}+\textit{i}t\right),\;\;\;\;\;\;\\
\label{GTformula}
&&G_T=-\frac{\textit{i}\pi }{2v_F}\int_{-\infty}^{\infty} dt
\frac{\sinh\left(\pi T t\right)}{\cosh^2\left(\pi T t\right) }\mathcal{T}\left(\frac{1}{2T}+\textit{i}t\right),\;\;\;\;\;\;\\
\label{kappaformula} 
&&K=\frac{ \pi^2 T}{v_F}\int_{-\infty}^{\infty}dt \frac{1}{\cosh^3\left(\pi T t\right)}\mathcal{T}\left(\frac{1}{2T}+\textit{i}t\right)-T\pi^2G.\;\;\;\;\;\; 
\end{eqnarray}
}

As follows from Eqs. (\ref{Gformula})--(\ref{kappaformula}), the thermoelectric transport properties of the considered system are completely defined by the $T$ matrix $\mathcal{T}$. For direct tunneling (dt), the leading term contributing to the $T$ matrix is proportional to the two-point finite-temperature Matsubara GF $\mathcal{G}_T(\tau)$, $\mathcal{T}_{dt}(\tau)\simeq  \lambda^2\mathcal{G}_T(\tau)$. In the inelastic cotunneling case (in), it is expressed via the four-point finite-temperature cSYK correlator $\mathcal{F}_T(\tau)$, $\mathcal{T}_{in}(\tau)\simeq  \lambda^4 \frac{T}{\sin\left(\pi T\tau \right)}\mathcal{F}_T(\tau)$ \cite{comment_orders}.

\paragraph*{Reparametrization modes.}
The Hamiltonian (\ref{Hsyk}) is invariant under $U(1)$ and $SL(2,\mathbb{R})$ symmetry reparametrizations \cite{Sachdev2015}. It does not change after the transformation $c_i(\tau)\rightarrow e^{-\textit{i}\phi(\tau)}\left[\dot{h}(\tau) \right]^{1/4}c_i(h(\tau))$,
where $h(\tau)$ is a monotonic time reparametrization with winding number $1$, and $\phi$ is a phase fluctuation with possibly arbitrary integer winding number \cite{Altland2019, Gu2019}.

These symmetries are broken by both the time derivative in the action and by additional terms in Eq.(\ref{FullH}). This leads to an effective action associated with energy costs of $\phi(\tau)$ and $h(\tau)$ fluctuations. As is shown in \cite{Altland2019}, the effective action $S_{\rm eff}$ splits into two independent parts for the corresponding fluctuations up to $1/N$ corrections, $S_{\rm eff}=S_{h}+S_{\phi}$, where
$ S_{h}=-m\int d\tau Sch[h(\tau),\tau];
S_{\phi}=16m\int d\tau \left[\phi^{\prime}(\tau)\right]^2$.
$Sch(h,\tau)\equiv (h^{\prime \prime}/h^{\prime})^{\prime}-\frac{1}{2}(h^{\prime \prime}/h^{\prime})^2$ is the Schwarzian operator, $m=\frac{N\ln N}{64J}\sqrt{\frac{\cos 2\theta}{2\pi}}$ is the effective mass appearing during the renormalization procedure \cite{Altland2019}, and $\theta$ is related to the average hole occupation $Q$ as $Q\equiv \frac{\langle n\rangle}{N}=\frac{1}{2}-\frac{\theta}{\pi}-\frac{\sin(2\theta)}{4}$ \cite{Sachdev2015, comment_conformal}.

\paragraph*{Conformal regime. }
Here we consider tunneling in the conformal case of the cSYK model, valid at temperatures $T^*\ll T\ll J$. The tunneling in the system is a sum of the elastic and inelastic processes.

Direct tunneling dominates at high temperatures $T\gg E_C$ \cite{Altland2019, Can2019}. The $T$ matrix for elastic processes in the leading order is $\mathcal{T}(\tau)={\color{black}\lambda^2} G^c(\tau)D(\tau)$. The conformal cSYK GF $G^c(\tau)$ is
\beq \label{Gbasic}  G^c(\tau)=-C{\rm sgn}(\tau)\sin\left(\pi/4-{\rm sgn}(\tau)\theta\right)\left(\frac{TJ}{\sin(\pi T|\tau|)}\right)^{1/2}, 
\eeq
$C=\left[(8/\pi)\cos(2\theta)\right]^{-1/4}$, and the spectral asymmetry parameter $\mathcal{E}$ defines $\theta$ as $e^{2\pi\mathcal{E}}=\tan(\theta+\pi/4)$ \cite{Gu2019, comment_sign}.
The two-point Coulomb correlator $D(\tau)$ reads
\beq \label{DCoulomb} D(\tau)=\frac{\mathcal{\theta}_3\left(-\textit{i}E_C\tau-\textit{i}\mathcal{E}\pi,e^{-\frac{E_C}{T}}\right)}{\mathcal{\theta}_3\left(-\textit{i}\mathcal{E}\pi,e^{-\frac{E_C}{T}}\right)}e^{-E_C|\tau |}, \eeq
where $\mathcal{\theta}_3(\bullet,\ast)$ is the Jacobi theta function \cite{comment_theta} (see Supplemental Material \cite{supplemental}).      

In this regime, the electric conductance $G$ is suppressed with $T$ by the Arrhenius exponential factor $G$$\sim $$e^{-\frac{E_C}{T}}$ at 
$T$$\ll $$E_C$, while $G$$\sim $$\frac{1}{\sqrt{T}}$ at $T$$\gg $$E_C$ \cite{Altland2019, Efetov2003, Fazio1991, Kamenev1996, Sedlmayr2006, Kiselev2012}. 
Thermopower $\mathcal{S}$ in this regime grows at small temperatures, while it saturates to a constant proportional to the spectral asymmetry parameter $\mathcal{E}$ at large $T$ \cite{Sachdev2015}. In the pure SYK system without charging energy of the dot, the Lorenz ratio $L=\lim_{T\rightarrow 0}\frac{\kappa}{T G}$ at zero temperature is modified in accordance with \cite{Davison2017}. The Wiedemann-Franz law breaks down at finite temperatures if the Coulomb blockade effect is present in the system ($T\ll E_C$), similar to \cite{Kubala2008}.

An inelastic cotunneling process dominates at low temperatures $T^*\ll T \ll E_C$, where the direct tunneling is exponentially suppressed by the Coulomb blockade. 
The leading term in the inelastic component of the $T$ matrix $\mathcal{T}_{in}(\tau)$ consists of the two-point free fermionic correlator, four-point SYK $F_{\rm SYK}$, and Coulomb $F_C$ correlators: $\mathcal{F}_T(\tau)\simeq F_{\rm SYK}(\tau)F_{C}(\tau)$ (see Supplemental Material \cite{supplemental}).  As shown in \cite{Maldacena2016}, the leading term in the $1/N$ orders of the SYK four-point function in the conformal limit is a factorization of the two-point functions given by Eq. (\ref{Gbasic}),
$ F_{\rm SYK}(\tau)\simeq G^c(\tau)G^c(-\tau)$ .   

Evaluation of Eq. (\ref{Gformula}) in this case gives that the electric conductance $G$ is linear in temperature $G\sim T$.

The electric conductance $G$ of both the elastic and inelastic processes was analyzed in \cite{Altland2019, Kruchkov2019} (see Supplemental Material \cite{supplemental}). In the conformal regime, $G$ is linear in temperature in the $T\ll E_C$ limit, but the direct tunneling quickly becomes dominant with increase of temperature \cite{comment_asymmetry}.

The thermoelectric coefficient $G_T$ is exponentially small in the $T\ll E_C$ limit for both the elastic and inelastic processes; the leading contribution is elastic, and gives $G_T\sim e^{-\frac{E_C}{T}}$.

\paragraph*{Schwarzian regime.}
In this section, we study the transport properties away from the conformal regime. This case is realized at $T \ll T^*$. 
The physics of the direct tunneling is defined by the two-point Green's function.
The two-point Coulomb correlator in this case is again given by Eq.(\ref{DCoulomb}). In this region of parameters, the SYK GF is strongly renormalized by the soft mode $h(\tau)$. The exact form of the temperature-dependent GF in time representation was found in \cite{Mertens2017}. In the low-temperature limit, the renormalized GF is 
\beq \label{GRenEl} G^r(\tau)\rightarrow -{\rm sgn}(\tau)\frac{\beta^{3/2}}{(4\pi)^{\frac{1}{4}}}\frac{\Gamma^4(\frac{1}{4})}{\pi}\frac{m e^{-\frac{\pi^2 }{\beta m}}}{|\tau|^{3/2}(\beta-|\tau|)^{3/2}}, \eeq
Here, $\beta $$\equiv$$ $$1/T$.
The contribution from the direct tunneling is strongly suppressed by the Arrhenius exponent $e^{-\frac{E_C}{T}}$ from the Coulomb correlator.

In the inelastic cotunneling case, the four-point SYK correlator is renormalized by the $h(\tau)$ modes. Following \cite{Mertens2017, Bagrets2017}, one can express this correlator $F_{\rm SYK}(\tau)=\left\langle G_{\tau}[h]G_{-\tau}[h] \right\rangle_h$ in the limit of low temperature as
\beq F^r(\tau)\rightarrow \frac{\beta^{\frac{3}{2}}\pi m^{\frac{1}{2}}e^{-\frac{\pi^2 }{\beta m}}}{2^{\frac{3}{2}}|\tau|^{\frac{3}{2}}(\beta-|\tau|)^{\frac{3}{2}}}. \eeq

\begin{figure} [t!]
\centerline{\includegraphics [width=75mm] {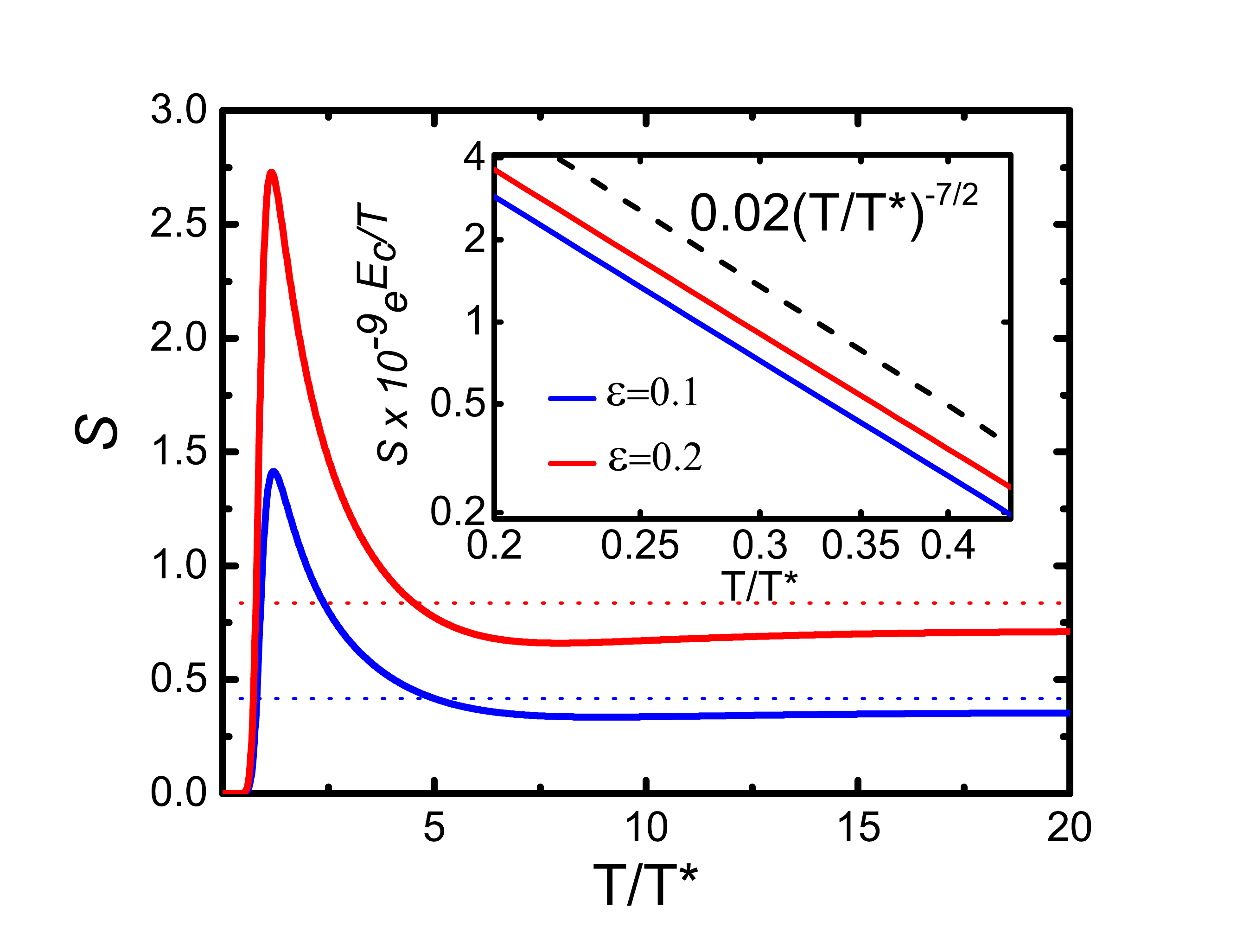}} \vspace{-1mm}
\caption{Thermopower $\mathcal{S}$ accounting for direct tunneling and inelastic cotunneling in the conformal regime for  $\mathcal{E}=0.1$ (blue line) and $\mathcal{E}=0.2$ (red line), $N=50$, $E_C/T^*=10$, $(\lambda/T^*)^2=0.03$. Dotted lines are asymptotic limits $\frac{4\pi}{3}\mathcal{E}$ of thermopower for corresponding values of $\mathcal{E}$. Inset: $Ln$$-$$ln$ plot for $\mathcal{S}$ multiplied by $e^{-\frac{E_C}{T}}$ in the Schwarzian regime for $\mathcal{E}=0.1$ (blue line) and $\mathcal{E}=0.2$ (red line). Black dotted line is $0.02(T/T^{\ast})^{-7/2}$. Inset demonstrates the $(T/T^{\ast})^{-7/2}e^{-\frac{E_C}{T}}$ law for thermopower \cite{comment_correction}.}
\label{fig.Thermopower}
\end{figure}

This correlator defines the scaling law of the electric conductance as 
$G\sim T^{3/2}$. The four-point Coulomb correlator does not depend on $T$ and 
$\mathcal{E}$ in the leading order, so the electric conductance $G$ retains the power-law scaling. In contrast, the thermoelectric coefficient $G_T$ is always exponentially suppressed at $T$$\ll $$E_C$. It follows from the symmetries of Eq.(\ref{GTformula}), since all nonvanishing contributions are exponentially small. In the Schwarzian regime, the leading contribution is elastic, $G_T\sim T^{-2}e^{-\frac{E_C}{T}}$ (same as in the conformal regime). This fact changes the thermopower $\mathcal{S}$ below $E_C$ both in the conformal and Schwarzian regimes. The resulting thermopower is depicted in Fig.\ref{fig.Thermopower}. It saturates to 
$\mathcal{S}$$=$$\frac{4\pi}{3}\mathcal{E}$ at $T$$\gg $$E_C$, approaching this limit as $\sim\frac{1}{\sqrt{T}}$, but it is exponentially suppressed at $T$$\ll $$E_C$, so it reaches zero instead of having divergence, which is expected if one considers only the direct tunneling in the system. 
\begin{equation} \label{thermopower}
\mathcal{S} \sim \begin{cases}
(T^{\ast}/T)^3e^{-\frac{E_C}{T}}, & T^*\ll T \ll E_C \\
(T^{\ast}/T)^{\frac{7}{2}}e^{-\frac{E_C}{T}}, & T\ll T^*
\end{cases}.
\end{equation}

The same analysis is applicable to the Peltier coefficient $\Pi$ due to its connection to thermopower, $\Pi$$=T$$ \mathcal{S}$ \cite{Zlatic2014}. Both the thermopower and the Peltier coefficient are antisymmetric in the spectral asymmetry parameter $\mathcal{E}$ and exponentially suppressed at low temperature by the Arrhenius exponent. The exact dependence of $\mathcal{S}$ on $\mathcal{E}$ is shown in the Supplemental Material \cite{supplemental}.



\begin{figure} [t!]
\centerline{\includegraphics [width=75mm] {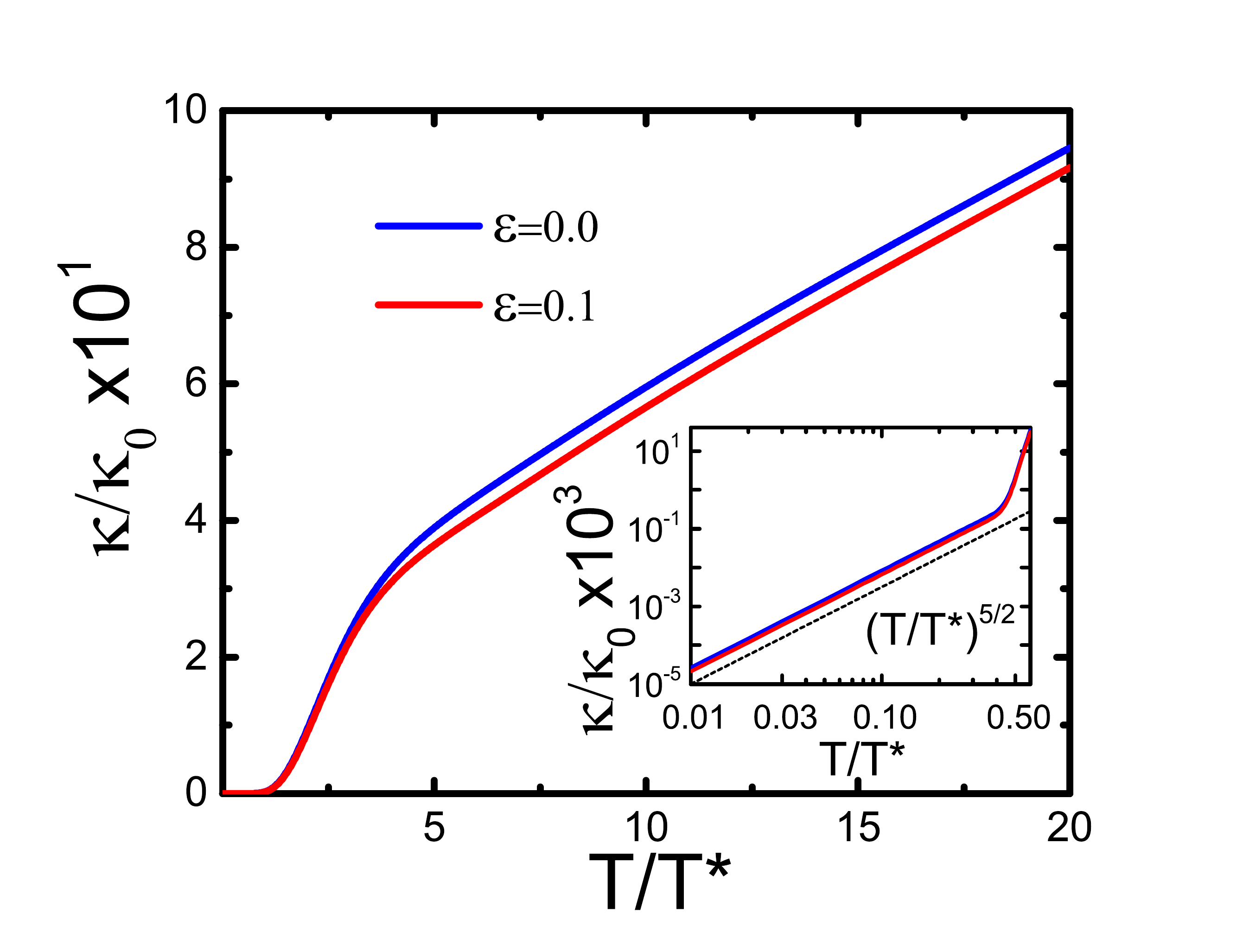}} \vspace{-1mm}
\caption{Thermal conductance $\kappa$ accounting for elastic and inelastic cotunneling in the conformal regime, $N=50$, $E_C/T^*=10$, $(\lambda/T^*)^2=0.03$,  $\kappa_0=T^{*\,3}/v_F$. The lines correspond to $\mathcal{E}=0$ (blue line) and $\mathcal{E}=0.1$ (red line). Inset: $Ln$$-$$ln$ plot for thermal conductance $\kappa$ in the Schwarzian regime for the same values of $\mathcal{E}$.}
\label{fig.kappa}
\end{figure}

\paragraph*{Thermal conductance. } As follows from Eq.(\ref{kappaformula}) and our analysis of the charge conductance and thermoelectric coefficient above, at temperatures below $T\ll E_C$, both in the conformal and Schwarzian regimes, the contribution to the thermal conductance $\kappa$ proportional to $G_T^2$ is exponentially suppressed, and the temperature dependence of $\kappa$ has the same scaling in the leading order as $T G$ and stems form the inelastic cotunneling. The temperature dependence of $\kappa$ is shown in Fig. \ref{fig.kappa}.

\paragraph*{Lorenz ratio.}
The authors of \cite{Davison2017, Kruchkov2019} have demonstrated that the Wiedemann-Franz law and the Lorenz ratio \cite{Zlatic2014, Benenti2017, Karki2020s} are violated in the conformal regime of the cSYK model, $L_{\rm SYK}=\frac{\pi^2}{5}$. Accounting for the inelastic cotunneling contribution to the transport coefficients in this regime, we come to the Lorenz ratio
\beq \nn L_{in}=\lim_{T\rightarrow 0}\frac{\kappa (T,\mathcal{E}=0)}{T G(T,\mathcal{E}=0)}=\frac{\pi^2}{2}. \eeq  

Considering the same entity in the Schwarzian case, we come to the Lorenz ratio $L_{in}\simeq 0.52\pi^2$.

\paragraph*{Discussion. }
We considered the effects of inelastic cotunneling in the thermoelectric transport of the cSYK model in conformal and Schwarzian limits. We demonstrated that even in the weak-tunneling limit, the related tunneling process gives the leading order contribution to electric conductance $G$, and thermal conductance $\kappa$. The thermoelectric coefficient $G_T$ is exponentially suppressed in both the elastic and inelastic cotunneling regimes at temperatures below $E_C$, while the electric and thermal conductances retain the power-law behavior. This selective suppression of the thermoelectric coefficient appears as a consequence of the particle-hole symmetry breaking. The diagonal transport coefficients are not sensitive to small particle-hole asymmetry and have finite values in the particle-hole symmetric point, while the off-diagonal coefficient $G_T$ is nonzero only when this symmetry is broken. It is proportional to the asymmetry parameter $\mathcal{E}$ (at $\mathcal{E}\ll 1$). All the terms contributing to the two-point and four-point Coulomb correlators that contain $\mathcal{E}$ also contain some power of the Arrhenius exponent, which leads to the reported low-$T$ suppression of $G_T$. On the contrary, the diagonal transport coefficients stem at low temperatures predominantly from the $\mathcal{E}$-independent term of the four-point correlator, which does not have the exponential suppression at low temperatures, so they exhibit a power-law behavior in $T$. As a direct consequence of this thermoelectric Coulomb blockade, thermopower $\mathcal{S}$ is exponentially suppressed as well. In general, the importance of inelastic cotunneling in the conductance peaks $\left \langle Q\right\rangle=N$ was underlined in \cite{Matveev2002, Khveshchenko2020L, Khveshchenko2020}, while in the conductance valleys $\left\langle Q\right\rangle=N+\frac{1}{2}$, the direct tunneling usually accounts for all relevant contributions \cite{Beenakker1992}. However, in the Schwarzian regime, the effective charging energy always has nonzero value, so the inelastic contribution is crucial here, resulting in the exponential suppression of $\mathcal{S}$.  The selective Coulomb blockade of different transport coefficients, namely, electric and thermal conductances, were recently observed experimentally in the context of the heat Coulomb blockade \cite{Sivre2017}. 

The Kelvin formula for thermopower, rigorous for the cSYK model at $T$$\gg $$E_C$, is not applicable when the Coulomb blockade effects cannot be neglected. This discrepancy arises due to the leading role of the inelastic processes, so the transport coefficients now have a different energy dependence from the DOS \cite{Mravlje2016}.

Even when the transport coefficients of the cSYK model stem from the inelastic cotunneling contribution, the Lorenz ratio of the cSYK model is not sensitive to the renormalization of the model by reparametrization modes and the Coulomb blockade effects. The similar finite-temperature relation (the Wiedemann-Franz law) is violated due to the Coulomb blockade effect. 

\paragraph*{Conclusion.} In this {\it Letter}, we analyzed the temperature behavior of the charge and heat transport coefficients in the Schwarzian regime of the cSYK model.  We showed that both electric and thermal conductance obey power a law in the temperature behavior characteristic for non-Fermi liquid regimes while Seebeck and Peltier coefficients are exponentially suppressed. The leading contribution to the transport coefficients in this regime is given by the inelastic processes. We suggest that future works test the theoretical predictions in quantum transport experiments in semiconductor nanostructures.


\paragraph*{Acknowledgements.} The work of M.K. was supported in part by the National Science Foundation under Grant No.NSF PHY-1748958 and conducted within the framework of the Trieste Institute for Theoretical Quantum Technologies (TQT). The authors are grateful to D. B. Karki for discussions on the Wiedemann-Franz law and transport coefficients.

\bibliography{refSYK1}

\begin{thebibliography}{85}
\bibitem [{com({\natexlab{b}})}]{comment_Ec}%
  \BibitemOpen
  \href@noop {} {\emph {\bibinfo {title} {{\rm In a realistic quantum dot, the
  charging energy $E^{(0)}_C=\frac{e^2}{2C_0}$, where $C_0$ is the dot's
  capacitance, scales with the dot linear size $l$ as $E^{(0)}_C\sim l^{-1}$,
  while the mean level spacing $\delta\sim \frac{E^{(0)}_C}{k_F l}$, with $k_F
  l\gg 1$. Here, $k_F$ is the Fermi wave vector \cite{Aleiner2002}. The
  separation between the lead and the dot is $d$; the electrons in the lead are
  screened from the dot by the Yukawa potential $U(r)\sim
  \frac{1}{r}e^{-r/r_d}$ with $r_d\gg d$. }}}}\BibitemShut {NoStop}%
\end{thebibliography}
\onecolumngrid
\newpage
\begin{center}{\large\bf Quantum thermal transport in the charged Sachdev-Ye-Kitaev model:\\ Thermoelectric Coulomb blockade\\\mbox{}\\}
{\large\bf Supplemental Material}

\end{center}
\begin{center}{Andrei I. Pavlov{$^1$}, Mikhail N. Kiselev$^1$}
\end{center}
\begin{center}{\fontsize{9}{11}\textit{ $^1$The Abdus Salam International Centre for Theoretical Physics (ICTP), Strada Costiera 11, I-34151 Trieste, Italy}}
\end{center}
\setcounter{equation}{0}
\setcounter{figure}{0}
\setcounter{table}{0}
\setcounter{section}{1}
\makeatletter\@addtoreset{equation}{section}
\@addtoreset{equation}{subsection} \makeatother
\renewcommand{\theequation}{\thesection.\arabic{equation}}
\renewcommand{\thesection}{S\arabic{section}}
\renewcommand{\thesubsection}{(\alph{subsection})}
\renewcommand\thefigure{S\arabic{figure}}
\renewcommand\bibnumfmt[1]{[#1]}
\renewcommand\citenumfont[1]{#1}
\numberwithin{equation}{section}
\counterwithin*{equation}{section}
\counterwithin*{equation}{subsection}

We  adopt the notations and definitions of the main text and use
the numeration of equations and references of the \textit{Letter}.

\section{Derivation of $G$, $G_T$ and $\kappa$}

In accordance with Eq.(\ref{currentI}), the transport coefficients can be found by taking derivatives of Eq.(\ref{FGL}) by voltage at uniform temperature (coefficient $G$) and by temperature at zero voltage (coefficient $G_T$). Here we perform this procedure explicitly.

Let us start with the conductance $G$. It is defined as
\beq  G= \left. \frac{I}{\Delta V}\right\vert_{\Delta V=0}= \left. 2\pi\int_{-\infty}^{\infty}d\varepsilon \rho_a(\epsilon)\rho_c(\epsilon)\partial_{\Delta V}\left(f(\epsilon +\Delta V, T+\Delta T)-f(\epsilon, T) \right)\right\vert_{ \Delta V=0}. \eeq

$\rho_a(\epsilon)=(2\pi v_F)^{-1}$ (see the main text), so this expression is further simplified as 
\beq \label{Gexpr} G=-\frac{1}{4v_F T}\int_{-\infty}^{\infty}d\epsilon \frac{\rho_c(\epsilon)}{\cosh^2\left(\frac{\epsilon}{2T}\right)}. \eeq
It was shown in \cite{Matveev2002} that the density of states $\rho_c(\epsilon)$ can be expressed in a general form via the T-matrix $\mathcal{T}$ in the real time representation
\beq \label{DoSTime} \rho_c(\epsilon)=-\frac{1}{\pi}\cosh\left(\frac{\epsilon}{2T}\right)\int_{-\infty}^{\infty}dt\mathcal{T}\left(\frac{1}{2T}+\textit{i}t\right)e^{\textit{i}\epsilon t}. \eeq 

Combining Eqs.(\ref{Gexpr}) and (\ref{DoSTime}), we have
\beq G=\frac{1}{4\pi v_F T}\int_{-\infty}^{\infty}dt\mathcal{T}\left(\frac{1}{2T}+\textit{i}t\right)\int_{-\infty}^{\infty}d\epsilon\frac{e^{\textit{i}\epsilon t}}{\cosh\left(\frac{\epsilon}{2T}\right)}=\frac{1}{2v_F}\int_{-\infty}^{\infty}\frac{dt}{\cosh\left(\pi Tt\right)}\mathcal{T}\left(\frac{1}{2T}+\textit{i}t\right). \eeq

$G_T$ is obtained in the same way. 
\beq \label{GTexpr} G_T=\left. \frac{I}{\Delta T}\right\vert_{\Delta T=0}=\left. 2\pi \int_{-\infty}^{\infty}d\varepsilon \rho_a(\epsilon)\rho_c(\epsilon)\partial_{\Delta T}\left(f(\epsilon, T+\Delta T)-f(\epsilon, T) \right)\right\vert_{\Delta T=0}. \eeq
Plugging DOS Eq.(\ref{DoSTime}) in Eq.(\ref{GTexpr}), we obtain
\beq G_T=\frac{\textit{i} }{4v_F \pi T^2}\int_{-\infty}^{\infty}dt \mathcal{T}(\frac{1}{2T}+\textit{i}t)\frac{\partial}{\partial t}\frac{2\pi T}{\cosh \left( \pi T t\right)}=-\frac{\textit{i}\pi }{2v_F}\int_{-\infty}^{\infty}dt\frac{\sinh\left(\pi Tt\right)}{\cosh^2\left(\pi Tt\right)}\mathcal{T}\left(\frac{1}{2T}+\textit{i}t\right). \eeq
By treating the thermal current in the similar way, we get the coefficient $K$ and the thermal conductance $\kappa$.
\bea K=-\frac{1}{4v_FT}\int_{-\infty}^{\infty}dt\mathcal{T}\left(\frac{1}{2T}+\textit{i}t\right)\frac{\partial^2}{\partial t^2}\frac{1}{\cosh\left( \pi T t\right)}=
-\frac{ \pi^2 T}{2v_F}\int_{-\infty}^{\infty}dt\mathcal{T}\left(\frac{1}{2T}+\textit{i}t\right)\left\{\frac{1}{\cosh \left(\pi T t\right)}-\frac{2}{\cosh^3\left(\pi T t\right)} \right\}
,\;\;\;\;\;\;\;\eea
here we used the identity $\sinh^2 x\equiv\cosh^2 x-1$. 
\beq \kappa =K-\frac{1}{T}\frac{G_T^2}{G}=\frac{\pi^2 T}{v_F}\int _{-\infty}^{\infty}dt\mathcal{T}\left(\frac{1}{2T}+\textit{i}t\right)\frac{1}{\cosh^3\left(\pi T t\right)}-T\pi^2G-\frac{1}{T}\frac{G_T^2}{G}
 \eeq
These expressions are Eqs. (\ref{Gformula}-\ref{kappaformula}) from the main text.

\section{Correlators} \label{correlators}
\setcounter{section}{2}
\setcounter{equation}{0}
\subsection{Two-point Coulomb correlator $D(\tau_1, \tau_2)$}
Let us reproduce here the derivation on then so-called Coulomb boson correlator arising due to averaging of the $U(1)$ gauge field:
\beq \label{Dt1t2} D(\tau_1,\tau_2)=\left\langle e^{-\textit{i}\phi(\tau_1)}e^{\textit{i}\phi(\tau_2)} \right\rangle_{\phi}. \eeq
The Green's function antiperiodicity condition $G(\frac{\beta}{2})=-G(-\frac{\beta}{2})$ imposes that $\phi(\frac{\beta}{2})=\phi(-\frac{\beta}{2})+-2\pi \textit{i}\mathcal{E}+2\pi W $, where $\mathcal{E}$ is the spectral asymmetry, $W$ in the winding number. We decompose the $\phi$ field such that $\phi(\tau)=\eta(\tau)+2\pi (W-\textit{i}\mathcal{E})T\tau$ introducing a periodic function $\eta(\tau)$: $\eta(\frac{\beta}{2})=\eta(-\frac{\beta}{2})$. The correlator (\ref{Dt1t2}) now reads
\bea D(\tau_1,\tau_2)=\frac{1}{Z_C}\sum_{W\in Z}\int D[\eta]e^{-\textit{i}\eta(\tau_1)}e^{\textit{i}\eta(\tau_2)}e^{-2\pi\textit{i}(W-\textit{i}\mathcal{E})T(\tau_1-\tau_2)}e^{-\int d \tau \eta^{\prime}(\tau)\frac{1}{4E_C}\eta^{\prime}(\tau)-\pi^2T\frac{(W-\textit{i}\mathcal{E})^2}{E_C}}\\
=\left\langle e^{-\textit{i}\eta(\tau_1)}e^{\textit{i}\eta(\tau_2)} \right\rangle_{\eta} \left\langle e^{-2\pi\textit{i}(W-\textit{i}\mathcal{E})T(\tau_1-\tau_2)} \right\rangle_W,\eea
$Z_C$ is the partition sum which normalizes the correlator

We start with averaging over the $\eta$ fields (we use here the Fourier image $\eta_m=\frac{1}{\beta} \int_0^{\beta} d\tau \eta_{\tau}e^{\textit{i}\omega_m\tau}$, $\omega_m=2\pi T m$):
\beq \label{sum_eta} \left \langle ... \right\rangle_{\eta}=\int D[\eta]e^{\textit{i}\int_{\tau_1}^{\tau_2} d\tau\eta^{\prime}(\tau)}e^{-\int d \tau \eta^{\prime}(\tau)\frac{1}{4E_C}\eta^{\prime}(\tau)}=\int D[\eta]e^{\beta\sum_{m\neq 0}\left(\frac{\omega_m^2\eta_{-m}\eta_m}{4E_C}-\textit{i}\eta_mJ_{-m}^{\tau_1; \tau_2} \right)},  \eeq
where $J_m^{\tau_1,\tau_2}=e^{\textit{i}\omega_m\tau_1}-e^{\textit{i}\omega_m\tau_2}$ is the Fourier image of $\delta(\tau-\tau_1)-\delta(\tau-\tau_2)$. The resulting integral over $\eta$ is Gaussian, it gives
\beq \left \langle ... \right\rangle_{\eta}=e^{-\beta\sum_{m\neq 0}\frac{E_C}{\omega_m^2}J_{-m}^{\tau_1,\tau_2}J_m^{\tau_1,\tau_2}},\;\;\;\;\;\;\;\;
q J_{-m}^{\tau_1,\tau_2}J_m^{\tau_1,\tau_2}=2\left(1-e^{\textit{i}\omega_m(\tau_2-\tau_1)} \right) 
\eeq
 

This $\left\langle ... \right\rangle_{\eta}$ correlator was discussed in 
\cite{Fazio1991, Kamenev1996, Efetov2003,  Kiselev2012, Sedlmayr2006}. Using the Sommerfeld-Watson transformation for Eq.(\ref{sum_eta}), one obtains 
\beq \label{correlator_eta} \left\langle ... \right\rangle_{\eta}=e^{-E_C(|\tau_2-\tau_1|-\frac{(\tau_2-\tau_1)^2}{\beta})}. \eeq
The other part of the Coulomb correlator, namely contribution from the winding numbers was evaluated in \cite{Efetov2003}.
Using the Poisson formula
\beq \sum_{k=-\infty}^{\infty}e^{-\frac{a}{2}k^2+\textit{i}xk}=\sqrt{\frac{2\pi}{a}}\sum_{n=-\infty}^{\infty}e^{-\frac{1}{2a}(x-2\pi n)^2}, \eeq
one can evaluate the partition sum $Z_C$ and the two-point propagator $\langle e^{2\pi\textit{i}T(W-\textit{i}\mathcal{E})(\tau_2-\tau_1)}\rangle_W$.
\beq Z_C=\sum_We^{-\frac{4\pi^2T(W-\textit{i}\mathcal{E})^2}{4E_C}}=\sum_me^{-\frac{E_C}{T}m^2+2\pi m\mathcal{E}}, \eeq
\beq \label{correlator_W} \left\langle ... \right\rangle_W=\frac{1}{Z_C}\sum_me^{-\frac{\pi^2T(W-\textit{i}\mathcal{E})^2}{E_C}+2\pi\textit{i}T(W-\textit{i}\mathcal{E})(\tau_2-\tau_2)}=\frac{1}{Z_C}\sum_me^{-\beta E_C(m-(\tau_2-\tau_1)T)^2+2\pi m \mathcal{E}}.  \eeq
The correlators above are normalized by corresponding partition sums. 
Quadratic terms in the exponents of both correlators (\ref{correlator_eta}) and (\ref{correlator_W}) cancel each other, so the result is
\beq D(\tau_1,\tau_2)=\frac{e^{-E_C|\tau_2-\tau_1|}}{Z_C}\sum_{m=-\infty}^{\infty} e^{-2E_C m (\tau_2-\tau_1)-\beta E_Cm^2+2\pi m \mathcal{E}}. \eeq

\subsection{Four-point Coulomb correlator}
Now we consider the same procedure applied to the four point Coulomb correlator
\beq \label{D1234} F(\tau_1,\tau_2,\tau_3,\tau_4)=\left\langle e^{-\textit{i}\phi(\tau_1)}e^{\textit{i}\phi(\tau_2)}e^{-\textit{i}\phi(\tau_3)}e^{\textit{i}\phi(\tau_4)} \right\rangle_{\phi}. \eeq
The approach is analogous to the former (two-point) case.  Decomposing $\phi(\tau)$ field into the periodic field $\eta(\tau)$ and the winding number contribution, we get factorization of two propagators.

\beq \left\langle e^{\beta\sum_{m\neq 0} \omega_m\eta_m(\int_{\tau_1}^{\tau_2}d\tau e^{\textit{i}\omega_m\tau}+\int_{\tau_3}^{\tau_4}d\tilde{\tau} e^{\textit{i}\omega_m\tilde{\tau}})} \right\rangle_{\eta}=e^{-\beta\sum_{m\neq 0}\frac{E_C}{\omega_m^2}(J_{-m}^{\tau_1;\tau_2}+J_{-m}^{\tau_3;\tau_4})(J_{m}^{\tau_1;\tau_2}+J_{m}^{\tau_3;\tau_4})}. \eeq
\beq (J_{m}^{\tau_1;\tau_2}+J_{m}^{\tau_3;\tau_4})=2\left(1-e^{\textit{i}\omega_m\tau_{41}} \right)+2\left(1-e^{\textit{i}\omega_m\tau_{32}} \right)+2e^{\textit{i}\omega_m\tau_{31}}-2e^{\textit{i}\omega_m\tau_{21}}+2e^{\textit{i}\omega_m\tau_{42}}-2e^{\textit{i}\omega_m\tau_{43}}, \eeq
where we denoted $\tau_{ij}\equiv \tau_i-\tau_j$.
\beq \left\langle ... \right\rangle_{\eta}=e^{-E_C\left(|\tau_{41}|+|\tau_{32}|-|\tau_{31}|-|\tau_{42}|+|\tau_{21}|+|\tau_{43}| \right)+\frac{E_C}{\beta}\left(\tau_{41}^2+\tau_{32}^2-\tau_{31}^2-\tau_{42}^2+\tau_{21}^2+\tau_{43}^2 \right)} \eeq
Averaging over winding numbers, we get
\beq \left\langle e^{2\pi\textit{i}(W-\textit{i}\mathcal{E})T(\tau_{21}+\tau_{43})} \right\rangle_W=\frac{1}{Z_C}\sum_me^{-\beta E_C(m-(\tau_{21}+\tau_{43})T)^2+2\pi m \mathcal{E}}. \eeq
The quadratic terms in these two correlators cancel each other, so the resulting four-point function is
\beq \label{F_corr} F(\tau_1,\tau_2,\tau_3,\tau_4)=\frac{1}{Z_C}\sum_me^{-\frac{E_C}{T}m^2-E_C\left(|\tau_{41}|+|\tau_{32}|-|\tau_{31}|-|\tau_{42}|+|\tau_{21}|+|\tau_{43}| \right)+E_C m(\tau_{21}+\tau_{43})+2\pi m \mathcal{E}}. \eeq
Let us choose some particular  time ordering, for instance, $\tau_1>\tau_4>\tau_2>\tau_3$. In this case, the correlator becomes
\beq F(\tau_1,\tau_2,\tau_3,\tau_4)=\frac{1}{Z_C}\sum_me^{-\frac{E_C}{T}m^2+E_C\left(\tau_{14}+\tau_{23}\right)(2m-1)+2\pi m \mathcal{E}}. \eeq
The main contribution to this sum comes from $m=0$. This term was discussed in \cite{Altland2019} and reads
\beq \label{4pointm0}e^{-E_C\left(\tau_{14}+\tau_{23}\right)}. \eeq
Note that only certain time orderings are relevant for the inelastic co-tunneling process. Namely, $\tau_1,\tau_4>\tau_2,\tau_3$ or $\tau_2,\tau_3>\tau_1,\tau_4$ are relevant, while other orderings (e.g. $\tau_1,\tau_2>\tau_3,\tau_4$) correspond to two sequential direct tunnelings (\citep{Bagrets2017}). For all 8 possible relevant time orderings, (\ref{F_corr}) simplifies to
\beq F(\tau_1,\tau_2,\tau_3,\tau_4)=\frac{1}{Z_C}\sum_me^{-\frac{E_C}{T}m^2-E_C\left(|\tau_{41}|+|\tau_{32}| \right)+E_Cm(\tau_{41}+\tau_{23})+2\pi m \mathcal{E}}. \eeq

$G$ and $G_T$ in the small tunneling approximation are proportional to
\beq \label{int4} \int d^4\tau e^{-E_C\left(\tau_{14}+\tau_{23}\right)}\left\langle G_{\tau_1,\tau_2}[h]G_{\tau_3,\tau_4}[h] \right\rangle_h,  \eeq
so the contributions from times $\tau_4\neq \tau_1$ and $\tau_3\neq \tau_2$ are exponentially suppressed. This allows to approximate
\beq \left\langle G_{\tau_1,\tau_2}[h]G_{\tau_3,\tau_4}[h] \right\rangle_h\simeq \left\langle G_{\tau_1,\tau_2}[h]G_{\tau_2,\tau_1}[h] \right\rangle_h \eeq
in the integral above.

We are interested in terms with non-zero $m$, which are exponentially small at low temperatures ($T\ll E_C$) comparing to (\ref{4pointm0}), so we can consider only $m=\pm 1$ terms.
Integrating Eq.(\ref{int4}) for various time orderings, one obtains a time-independent constant in the leading order. This term is dominant for conductivity, but first non-vanishing contribution to thermal conductivity comes from next terms, proportional to $e^{-E_C\tau_{12}+2\pi \mathcal{E}}$ and $e^{E_C\tau_{12}-2\pi\mathcal{E}}.$

\section{Electric conductance $G$}
\begin{figure} [t!]
\centerline{\includegraphics [width=1.\linewidth] {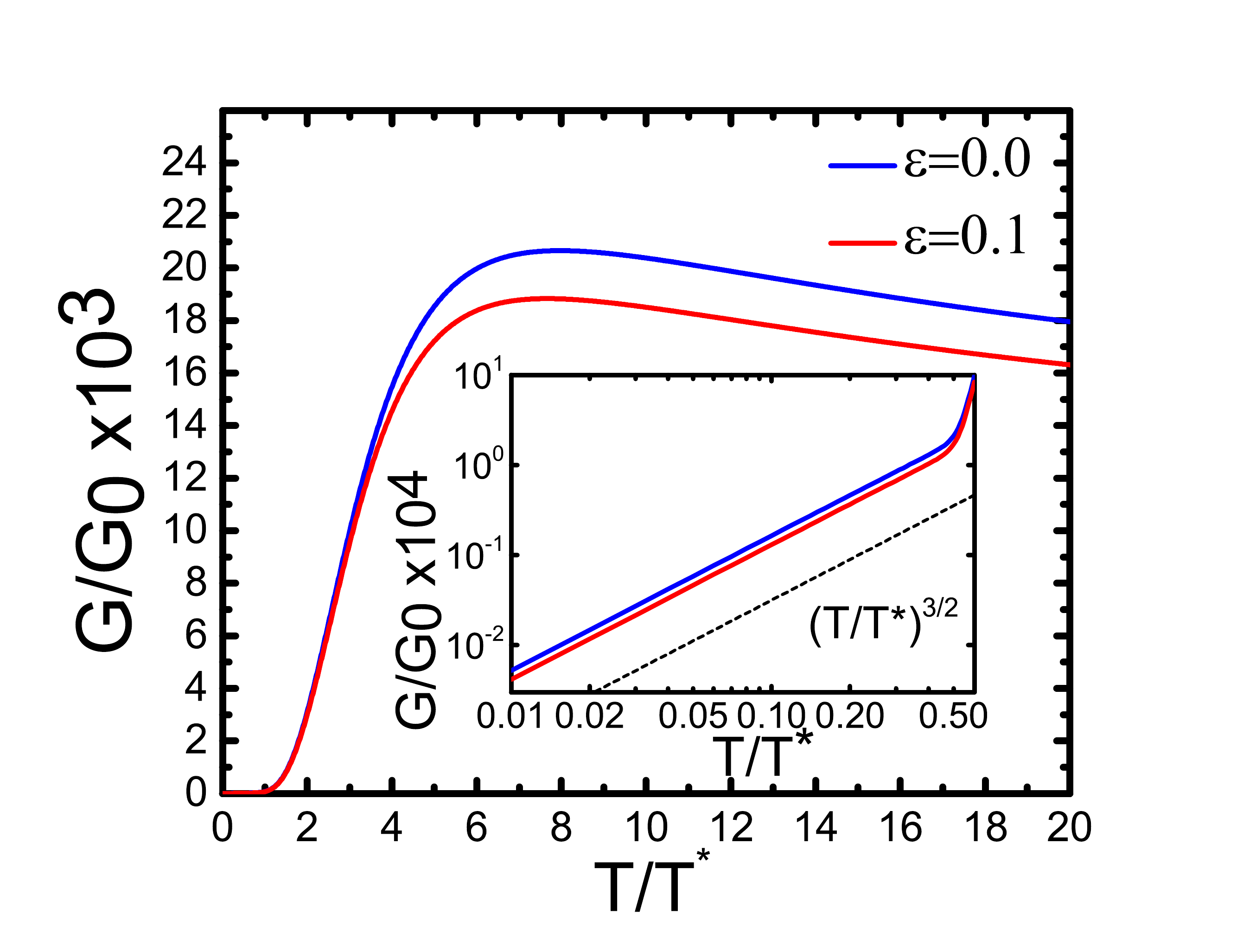}} \vspace{-1mm}
\caption{Electric conductance $G$ accounting for elastic and inelastic co-tunneling in the conformal regime, $E_C/T^*=10$, $N=50$, $G_0=\frac{(T^*)^2}{v_F}$, $(\lambda/T^{\ast})^2=0.03$. The lines correspond to $\mathcal{E}=0$ (blue) and $\mathcal{E}=0.1$ (red). Inset: $Ln$$-$$ln$ plot for electric conductance $G$ in the Schwarzian regime for the same values of $\mathcal{E}$.
}
\label{fig.Gfull}
\end{figure}

Here we provide our results for the electric conductance $G$ obtained by evaluation of Eq. (\ref{Gformula}). Fig. \ref{fig.Gfull} shows $G$ accounting for both elastic and inelasic processes. In the conformal regime, electric conductance scales as $G\sim \frac{1}{\sqrt{T}}$ for $T\gg E_C$ (this scaling comes from direct tunneling), while $G\sim T$ at $T^{\ast}\ll T\ll E_C$ (it stems from inelastic co-tunneling). The direct tunneling contribution becomes dominant with increase of temperature, so the intermediate region with dominant direct tunneling $G\sim e^{-E_C/T}$ is seen there at $T\simeq E_C$. The inset demonstrates the electric conductance in the Schwarzian regime of the theory $T\ll T^{\ast}\ll E_C$, here $G$ scales as $G\sim T^{3/2}$, this scaling stems from the inelastic co-tunneling. These results are in agreement with \cite{Altland2019, Kruchkov2019}. They are further used for evaluation of the thermopower $\mathcal{S}$ Eq. (\ref{thermopower}) of the main text.

\section{Thermopower $\mathcal{S}$ as a function of the spectral asymmetry parameter $\mathcal{E}$}
\begin{figure} [t!]
\centerline{\includegraphics [width=1.\linewidth] {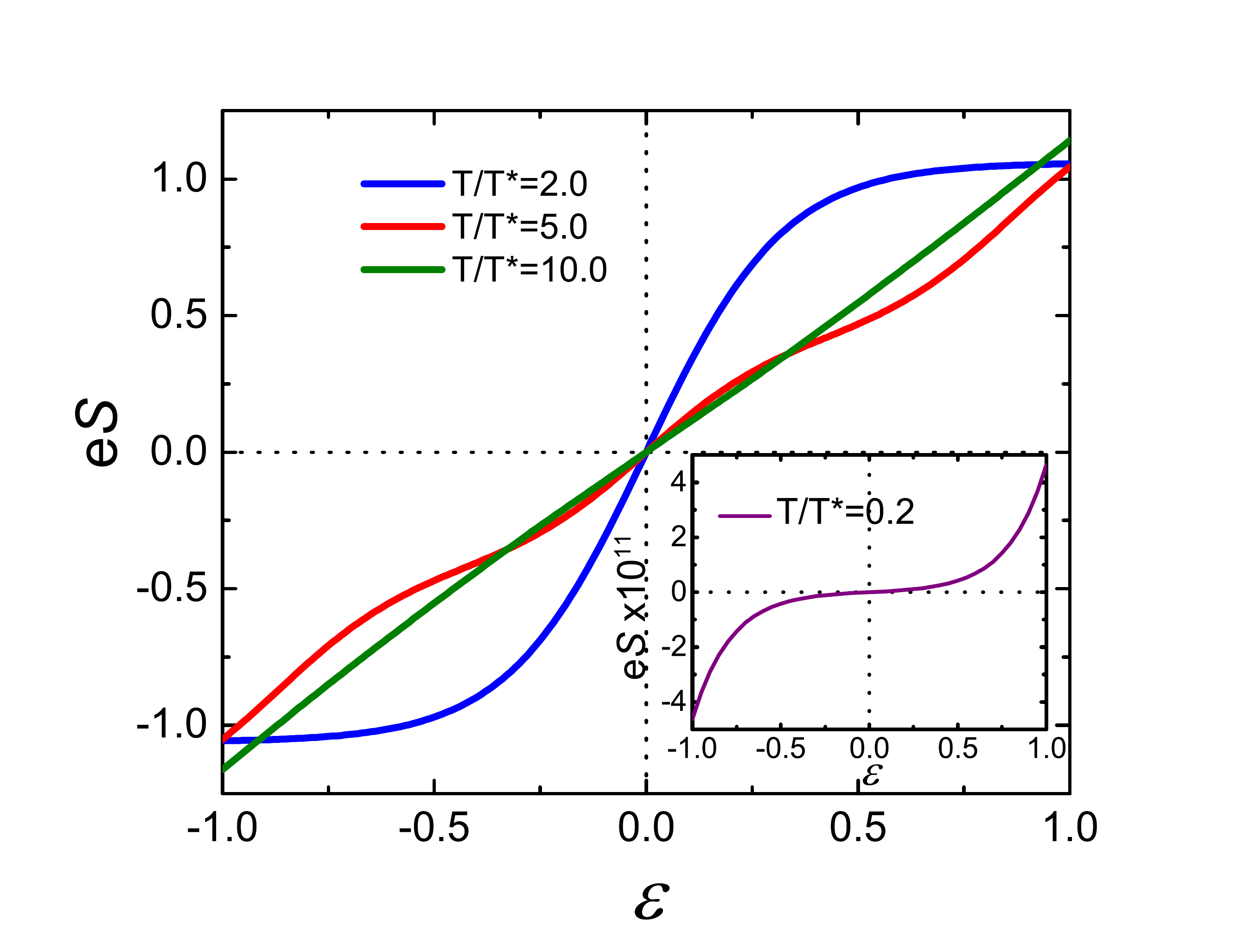}} \vspace{-1mm}
\caption{Thermopower $\mathcal{S}$ accountable for elastic
and inelastic co-tunneling in the conformal regime as a
function of $\mathcal{E}$, $N = 50$, $E_C/T^* = 10$, $(\lambda/T^*)^2 = 0.03$. $T/T^* = 2$ (blue line), $T/T^* = 5$ (red line), $T/T^* = 10$
(green line). Inset: Thermopower $\mathcal{S}$ in the Schwarzian regime, $T/T^* = 0.2$.
}
\label{fig.thermEps}
\end{figure}

As discussed in the main text, the thermopower $\mathcal{S}$ is antisymmetric in the spectral asymmetry parameter $\mathcal{E}$. $\mathcal{S}$ is linear in the leading order of $\mathcal{E}$ close to the particle-hole symmetric point ($\mathcal{E}\ll 1$). This behavior of the thermopower (multiplied on the electric charge $e$ to form dimensionless units) is plotted in Fig. \ref{fig.thermEps}. Note the scale of the inset showing the thermopower in the Schwarzian regime -  
$\mathcal{S}$ is exponentially suppressed by temperature in accordance with Eq. (\ref{thermopower}).

\end{document}